\documentclass[12pt]{article} \usepackage{amssymb,amsfonts}

\catcode`\@=11

\newif\if@fewtab\@fewtabtrue
{\count255=\time\divide\count255 by 60
\xdef\hourmin{\number\count255}
\multiply\count255 by-60\advance\count255 by\time
\xdef\hourmin{\hourmin:\ifnum\count255<10 0\fi\the\count255}}
\def\ps@draft{\let\@mkboth\@gobbletwo
    \def\@oddhead{}\def\@oddfoot{\hbox to 7 cm{\tiny \versionno
       \hfil}\hskip -7cm\hfil\rm\thepage \hfil}
    \def\@evenhead{}\let\@evenfoot\@oddfoot}
\def\aff           {affine Lie algebra}
\def\alg           {algebra}

\def\be            {\begin{equation}}
\def\bearl         {\begin{array}{l}}
\def\bearll        {\begin{array}{ll}}
\def\bearlll       {\begin{array}{lll}}

\def\bll           {\mbox{$b^{}_\ilabel$}} 
\def\bllch         {\mbox{$b^{\ch}_\ilabel$}}
\def\bllo          {\mbox{$b^{\jjj}_{\ilabel}$}}
\def\block         {\mbox{$\cal F$}}
\def\brev          {\breve}

\def\calw          {\mbox{$\Dot{\cal W}{.75}{.63}\;$}}

\def\Ccs           {\Dot C{.51}{.41}}

\def\cft           {conformal field theory}
\def\cfts          {conformal field theories}
\def\ch            {\Psi}
\def\chii          {\raisebox{.15em}{$\chi$}}
\def\chil          {\chii_\Lambda^{[J]}}

\def\chilP         {\chii_\LambdaP^{[J']}}
\def\chilP         {\chii_\LambdaP^{[J']}}
\def\chiO          {\raisebox{.15em}{$\brev\chi$}}
\def\chiOP         {\raisebox{.15em}{$\brev\chi{}'$}}
\def\cht           {{\tilde\Psi}}
\newcommand\ci[1]  {\cite{fusS#1}}

\def\cS            {{\Dot S{.75}{.41}}}

\def\cT            {{\Dot T{.75}{.46}}}

\def\cvira         {coset Virasoro algebra}

\def\dl            {\mathbb }
\newcommand\Dot[3] {{#1}\hspace{-#3em}\raisebox{#2em}
                   {$\scriptscriptstyle\bullet$}}

\def\ee            {\end{equation}}
\def\eE            {{\rm e}}
\def\eear          {\end{array}}

\def\emb           {\!\hookrightarrow\!}

\def\eq            {\,{=}\,}
\newcommand\erf[1]{(\ref{#1})}

\newcommand\fcft[3]{{{#1}^{\mskip-#3 mu\raise #2 pt\hbox{$\scriptstyle\circ$}}}}
\newcommand\Fa[2]  {F_a(#1,#2)}

\newcommand\fline  [1]{\vfill\noindent ------------------\\[1 mm]}
\def\fpr           {fixed point resolution}
\def\Fpr           {Fixed point resolution }
\newcommand\Frac[2]{\mbox{\large$\frac{#1}{#2}$}}

\def\futnot#1      {\ifnum\draftcontrol=1
                   \footnote{~{\sc internal footnote:} #1}\ \fi}
\def\futnote#1     {\footnote{~#1}\ }
\def\g             {{\liefont g}}
\def\G             {{G_{\rm id}}}
\def\gb            {\mbox{$\bar\g$}}
\def\gB            {{\bar\g}}
\def\gbP           {\mbox{$\gB'$}}

\def\gex           {{G_{\rm xt}}}
\def\Gex           {\mbox{$\gex$}}
\def\gg            {\mbox{$\liefont g$}}
\def\ggp           {$(\gb/\gbP)_\kV$}
\def\Gid           {\mbox{$G_{\rm id}$}}
\def\Gilabel       {G_{\ilabel}}
\def\Gilabels      {{G^{\phantom t}\!\!\!}^{\ \star}_{\,\ilabel}}

\def\Gjlabel       {G_{\jlabel}}

\def\gO            {\mbox{$\brev{\g}$}}
\def\gP            {\mbox{$\g'$}}
\def\gPO           {\brev\g'}
\def\gPP           {{\g'}}
\def\Gstab         {{G_\ilab}}
\def\h             {\mbox{$\Dot{\cal H}{.65}{.49}\;$}}

\def\hill          {\mbox{${\cal H}_\Lambda$}}

\def\hl            {\mbox{${\cal H}_\Lambda$}}

\def\hll           {\mbox{${\cal H}_{(\Lambda;\Lambda')}$}}
\def\hllch         {\mbox{${\cal H}^{\ch}_{(\Lambda;\Lambda')}$}}
\def\hlolo         {\mbox{${\cal H}_{(J\Lambda;J'\Lambda')}$}}
\def\hLP           {\mbox{${\cal H}_{\LambdaP}$}}
\newcommand\hsp[1] {\mbox{\hspace{#1 em}}}

\def\hy            {$\mbox{-\hspace{-.66 mm}-}$}
\def\Ia            {{\cal I}_a}
\def\Ib            {{\cal I}_b}
\def\ii            {{\rm i}}

\def\ilab          {{\Lambda;\Lambda'}}
\def\ilabel        {{(\Lambda;\Lambda')}}
\def\Ilabel        {{(\Lambda;\!\Lambda')}}
\def\ilabl         {{[\ilabel]}}
\def\Ilabl         {{[\Ilabel]}}

\def\iN            {\!\in\!}
\def\irmod         {irreducible module}
\def\jj            {{(J;J')}}
\def\JJ            {{(J;\!J')}}
\def\jjj           {{[J;J']}}

\def\jlabel        {{(\Mu;\Mu')}}

\def\kma           {Kac\hy Moo\-dy algebra}
\def\kv            {\mbox{$k_{}$}}
\def\kV            {{k}}
\def\kvo           {{\brev k}}

\long\def\labl#1   {\label{#1}\ee}

\def\LambdaOP      {{\brev\Lambda{}'}}

\def\LambdaP       {{\Lambda'}}
\def\Lc            {\Dot L{.74}{.48}\hsp{.25}}
\def\Lcs           {\Dot L{.51}{.39}\hsp{.1}}

\def\lie           {Lie algebra}

\def\liefont       {\mathfrak }

\def\lP            {{\Lambda'}}
\def\LP            {L'}

\def\modinv        {modular invarian}
\def\Mu            {\Upsilon}

\def\nstab         {{N_\ilab}}

\def\ocha          {twining character}
\def\olie          {orbit Lie algebra}

\def\om            {\omega}
\let\omchar=\ocha

\def\omP           {{\omega'}}
\def\omT           {\mbox{$\omega^\star$}}

\newcommand\omt[1] {{\omega^\star #1}}

\def\omTP          {{\omega'}^\star}
\newcommand\omtP[1]{{{\omega'}^\star #1}}
\def\one           {\mbox{\small $1\!\!$}1}

\def\qc            {q_{}^{\Lcs_0-\Ccs/24}}
\def\qdim          {quantum dimension}
\def\qfts          {quantum field theories}
\def\qg            {q_{}^{L_0-C/24}}

\long\def\query#1{\hskip 0pt{\vadjust{\everypar={}\small\vtop to 0pt{\hbox{}%
     \vskip -13pt\rlap{\hbox to 50.0pc{\hfil{\vtop{\hsize=8pc\tolerance=6000%
     \hfuzz=.5pc\rightskip=0pt plus 3em\noindent#1}}}}\vss}}}}%

\def\rep           {representation}
\def\Rep           {Representation}
\def\resp          {respectively}

\def\Sa            {{G_a}}
\def\Sb            {{G_b}}
\newcommand\Scc[4] {\cS^{}_{([(#1;#2)],\ch),([(#3;#4)],\cht)}}
\def\slz           {\mbox{SL(2,\Zet)}}

\def\So            {S^{[J]}}

\def\SoP           {S^{[J']}}

\def\smat          {S-matrix}

\def\tauc          {\mbox{$\Dot\tau{.49}{.46}_\jj$}}
\def\taug          {\tau^{}_{\!J}}
\def\taugP         {\tau^{}_{\!J'}}
\def\tbf           {twining branching function}

\let\tcha=\ocha
\def\trhl          {{\rm tr}^{}_{{\cal H}_{\Lambda}}}
\def\trhll         {{\rm tr}^{}_{{\cal H}_{\ilabel}}}
\def\trhlP         {{\rm tr}^{}_{{\cal H}_{\LambdaP}}}
\def\tS            {{\cal S}}
\def\tT            {{\cal T}}
\def\twodim        {two-di\-men\-si\-o\-nal}
\def\U             {{U}}

\newcommand\version[1] {\ifnum\draftcontrol=1 \typeout{}\typeout{#1}\typeout{}
                   \vskip3mm \centerline{\fbox{\tt DRAFT -- #1 -- \today}}  
                   \vskip3mm \fi}
\def\vir           {\mbox{$\Dot{\cal V}{.72}{.49}\,\liefont{ir}$}}
\def\vira          {Virasoro algebra}
\def\virg          {\mbox{${\cal V}\liefont{ir}_\g$}}
\def\virgP         {\mbox{${\cal V}\liefont{ir}_\gPP$}}

\def\wrt           {with respect to }
\def\wrtt          {with respect to the }
\def\wzwm          {WZW model}
\def\Wzwm          {Wess\hy Zu\-mi\-no\hy$\!$Witten (WZW) model}
\def\wzwt          {WZW theory}
\def\wzwts         {WZW theories}
\def\X             {\chii} 
\def\Xt            {{\cal X}}

\def\Zet           {${\dl Z}$}

\global\def\draftcontrol{0}
\catcode`\@=12

\begin{document}


\begin{flushright}  {~} \\[-23 mm]
{\sf DESY 96-251}\\ {\sf NIKHEF 96-032}\\{\sf UCB-PTH-96/57}, {\sf LBNL-39689}
\\[1 mm]{\sf hep-th/9612093}
\\[1 mm]{\sf December 1996} \end{flushright} \vskip 2mm\begin{center}\vskip12mm

{\Large\bf FIXED POINT RESOLUTION} \vskip.7em
{\Large\bf IN CONFORMAL FIELD THEORY\,$^{\#\!\!\!\!\#}$} \vskip 12mm
{\large J\"urgen Fuchs} $^1$\,, {\large Bert Schellekens} $^2$
\,\ and \,\ {\large Christoph Schweigert} $^3$
\\[9mm] {$^1$ \small DESY, Notkestra\ss e 85,\, D\,--\,22603~\,Hamburg}
\\[2mm] {$^2$ \small NIKHEF/FOM, Kruislaan 409,\, NL\,--\,1098 SJ~\,Amsterdam}
\\[2mm] {$^3$ \small Department of Physics, University of California,\,
Berkeley, CA 94720, USA}
\\ {\small and~~Theoretical Physics Group, Lawrence Berkeley National 
Laboratory,}\\{\small Berkeley, CA 94720, USA} \end{center}

\vskip 15mm

\begin{quote} {\bf Abstract.} \\
We summarize recent progress in the understanding of fixed point resolution 
for \cfts. Fixed points in both coset conformal field theories and 
non-diagonal modular invariants which describe simple current extensions of 
chiral algebras are investigated. A crucial r\^ole is played by
the mathematical structures of twining characters and orbit Lie algebras.
\end{quote}

\vfill {}\fline{} {\small
$^{\#\!\!\!\!\#}$~~Slightly extended version of a talk given by J.\ Fuchs at the
XXI International Colloquium on \\{}\mbox{$\,\ $}~~~Group Theoretical Methods 
in Physics (Goslar, Germany, July 1996)} \newpage

\section{Introduction}

Attempts to analyze \qfts\ beyond perturbation theory typically face enormous 
problems. For \twodim\ \cfts\ this task is certainly easier than for
higher-dimensional models, but even for the most extensively studied 
\Wzwm s based on simple Lie algebras a satisfactory field theory interpretation
is available only for a few of the known non-diagonal modular invariants.
Among the more severe obstacles there is the so-called {\em \fpr\/} problem, 
which arises when the orbits \wrt a symmetry have unequal sizes. (This kind 
of problem occurs in many other areas of physics and mathematics, too.) 

In this talk we report on recent progress in \fpr\ for 
various classes of \cfts\ which are based on \wzwts, namely 
coset theories \ci4 and simple current extensions of \wzwts\ \ci6.
These results imply a Verlinde formula for \wzwm s on
non-simply connected groups, and also led us to a conjecture 
concerning simple current extensions of arbitrary \cfts\ \ci6. Moreover,
they should play a r\^ole in the treatment of orbifolds of the type discussed 
in V.G.\ Kac's talk at this conference.
The mathematical structures emerging in our analysis of \fpr\ are those of
\tcha s and \olie s \cite{fusS3,furs}, which are explained in more detail in
the talk by C.\ Schweigert \cite{sfrs}.

\section{Coset \cfts}

The chiral algebra of a \wzwt\ is the semi-direct sum of
the \vira\ and an untwisted affine Kac\hy Moo\-dy \cite{K-M-KM} \lie\ \gg.
The well-established \rep\ theory of \aff s makes such theories amenable to
detailed study; they have therefore become prototypes and building blocks 
of \twodim\ \cft. The coset construction exploits the \rep\ theory of affine
algebras also for more complicated models. The basic idea is to consider 
embeddings $\gP\emb\g$, where both algebras are direct sums 
of untwisted affine \lie s and Heisenberg \alg s at some level(s) \kv. 
When the embedding of \gP\  into \gg\ is induced by an embedding 
$\gbP\emb\gb$ of the respective horizontal sub\alg s, then the difference of 
the two Virasoro algebras \virg\ and \virgP\ (which are obtained via the
affine Sugawara construction for \gg\ and \gP), i.e.\ the \lie\ with
generators $\Lc_m = L_m - \LP_m$, is again a \vira.

It is, however, an open problem whether for {\em any\/} embedding 
$\gbP\emb\gb$ this prescription of a \vira\ can be complemented in such a way
that one gets a consistent \cft\ -- called the {\em coset theory\/} and 
briefly denoted by `\,\ggp\,' -- and if so, whether that theory is unique.
To decide these questions, one must in particular construct
the (maximally extended) chiral algebra \calw\ of \ggp, as well as
the spectrum of the theory, i.e.\ tell which modules (\rep\ spaces) -- of 
the coset chiral \alg\ \calw, or at least of the \cvira\ \vir\ -- appear.

Concerning the spectrum, one first notes that even though \vir\ acts
on the chiral state space ${\cal H}_\g$ of the \wzwt\ for \gg, i.e.\ on the 
direct sum ${\cal H}_\g = \bigoplus_\Lambda \hill$ of all inequivalent unitary
irreducible highest weight modules $\hill$ of \gg\ at level \kv, 
${\cal H}_\g$ is {\em not\/} the state space \h\ of the coset theory. The 
crucial point is that \gP\ commutes with \vir, so that when
retaining the full space ${\cal H}_\g$, infinitely many spin zero fields,
namely all \gP-currents and their \gP-descendants, would be present.
This disaster is avoided by imposing the gauge principle that
${\cal H}_\g$-vectors differing only by the action of $\gP$
represent one and the same state in \h. Thus the elements of \h\ are
\gP-{\em orbits\/} of vectors in ${\cal H}_\g$ rather than individual
${\cal H}_\g$-vectors. An obvious ansatz is then to regard the 
`multiplicity spaces' $\hll$ that appear in the decomposition
  \be  \hl=\bigoplus\hsp{-.23}\raisebox{-.54em}{$\scriptstyle\LambdaP$}\
  \hll\otimes\hLP  \ee
of the \gg-modules $\hl$ into irreducible \gP-modules \hLP\ as the 
\ggp-modules, and correspondingly the branching functions 
$\bll$, which are defined by the decomposition 
  \be  \chii_\Lambda^{} = \sum_{\lP}\bll\, \chii^{}_{\lP} \ee
 of irreducible \gg-characters 
 \futnote{For notational simplicity here and below we only write the 
 \vir-specialized 
 characters. But all our results hold for the full characters. Note that
 the branching functions are indeed functions of $q$ rather than numbers.}
  \be  \chii_\Lambda:=\trhl(q_{}^{L_0-C/24})  \ee
 into irreducible \gP-characters $\chii_{\lP}:=\trhlP(q_{}^{L_0'-C'/24})$,
as the characters of the coset theory.

However, in general the redundancy is larger than the obvious gauge symmetry 
\gP, so that this construction is incomplete.
This manifests itself in selection rules, i.e.\ the branching functions
\bll\ for certain pairs $\Ilabel$ vanish, and equivalences,
i.e.\ multiplicity spaces $\hll$ for distinct pairs $\Ilabel$ are 
isomorphic modules. In particular the putative vacuum module
occurs several times. For the same reasons which force us to divide
out the \gP-action, this implies that a primary field of \ggp\
is not associated to an individual pair $\Ilabel$ of integrable highest 
weights, but rather to an
appropriate equivalence class $\ilabl$ of such pairs. This projection to
classes $\ilabl$ is known as {\em field identification}.

Both selection rules and field identification are conveniently described
by using the concept of the {\em identification group\/} $\G$ \cite{scya6}.
The elements $\JJ$ of $\G$ are primary fields
of the tensor product theory $\g\!\oplus\!(\gP)^*$ (the star indicates that one
must use the complex conjugate \slz\,-\rep) of unit \qdim\ (known as 
{\em simple currents\/}) and of integral spin; they act via the fusion product.
The selection rules are equivalent to the vanishing of the so-called
monodromy charges $Q_\jj$ of any allowed branching function with respect to all 
$\JJ\iN\G$.  (Here $Q_\jj(\Ilabel)=Q_J(\Lambda)-Q_{J'}(\Lambda')$, where 
$Q_J(\Lambda)$ is the combination $Q_J(\Lambda)=\Delta_\Lambda+\Delta_J-
\Delta_{J\star\Lambda}$ of conformal weights.) Moreover,
the equivalence classes in the field identification are precisely
the orbits of $\G$,
  \be  \ilabl = \{\, \jlabel \mid \Mu\eq J\Lambda,\;\Mu'\eq J'\Lambda' \
  {\rm for\ some}\ \JJ\iN\G \,\} \,. \ee
As long as all $\G$-orbits have a common length, taking one branching 
function out of each $\Gid$-orbit and combining them diagonally 
yields a modular invariant spectrum, and the modular \smat\ $\cS$ is given by
the restriction to orbits of the S-matrix $S\!\otimes\!(S')^*$
of $\g\!\oplus\!(\gP)^*$.

\section{Fixed point resolution in coset theories}

While the construction presented above works in many cases, in general
it is still insufficient. Indeed, as soon as {\em fixed points\/} -- that is, 
orbits $\Ilabl$ 
with non-maximal size (i.e.\ smaller than the size $|\G|$ of the orbit [(0;0)] 
that yields the vacuum module) -- are present, taking precisely one
representative out of each orbit does not give a modular invariant spectrum.
On the other hand, there does exist a modular invariant sesquilinear 
combination $Z$ of (non-zero) branching functions, namely
  \be  \tilde Z = \sum_{\scriptstyle \ilabl \atop\scriptstyle Q=0}
  |\Gstab| \cdot |\!\!\! \sum_{\jj\in\G/\Gstab} \!\!\!
  b_{\jj\,\ilabel}^{}\,|^2  \,, \labl Z
which is modular invariant. Here $\Gstab$ is the {\em stabilizer\/} of 
$\Ilabel$, i.e.\ the subgroup 
  \be  \Gstab=\{\JJ\iN\G\,|\,J\Lambda=\Lambda,\,J'\LambdaP=\LambdaP\} \ee
of those currents which leave $\Ilabel$ fixed; its order $|\Gstab|$ is the 
quotient of $N$ by the size $\nstab$ of the orbit $\ilabl$.

However, regarding the branching functions \bll\ appearing in \erf Z 
as irreducible \ggp-characters would 
lead to the conclusion that fractional multiplicities occur in $Z$, which does 
not allow for any sensible interpretation of $Z$ as a partition function.
The difficulty in interpreting $Z$ constitutes the {\em fixed point problem\/} 
of coset theories. This is indeed a severe problem, as it seems to prevent us 
from gaining control over the coset theory \ggp\ by completely understanding it 
in terms of the underlying \wzwts\ based on \gg\ and \gP. 
The solution to this problem (\ci4; earlier work is 
summarized in \cite{scya6}) tells us that \ggp\
is still fully controlled by the affine \alg s \gg\ and \gP, but that one needs
to implement additional novel structures \cite{fusS3,furs} associated to \kma s.

The basic observation is that the coset
modules $\hll$ for fixed points are {\em not\/} irreducible and hence
must be {\em resolved\/} into irreducible subspaces.
For the full \twodim\ theory obtained by combining its two chiral halves,
this means that not all states that one would naively expect are present,
i.e.\ an additional projection takes place on the space of states. 
To construct the submodules, we study the intertwiners between
isomorphic spaces $\hll$.
The simple currents $J$ and $J'$ act on the weights of \gg\ and \gP\ by maps 
$\omT$ and $\omTP$ (i.e.\ $J\Lambda\eq\omt(\Lambda)$, $J'\LambdaP
\eq\omtP(\LambdaP)$) which are induced by certain outer automorphisms
$\om$ and $\omP$ of \gg\ and \gP, \resp. 
\cite{sfrs}. Using these automorphisms we are able to construct a map
$\tauc\hsp{-.2}:\,\hll\!\to\!\hlolo$ which 
obeys $[\tauc,\Lc_m]=0$, i.e.\ intertwines the action of \vir.\,%
\futnote{For technical
reasons concerning the precise relation between $\om$ and $\omP$, the 
fixed point resolution works literally as sketched here only for so-called
`generalized diagonal coset theories'. But it is to be expected that
our ideas generalize to all coset theories except for very special classes.
The known exceptions (conformal embeddings and so-called maverick cosets)
all appear at levels $\kv\eq1$ or $\kv\eq2$.}

If $\Ilabel$ is fixed by $\JJ$, then $\tauc$ is an {\em endo\/}morphism, so 
that we can decompose $\hll$ as 
  \be  \hll=\bigoplus_\ch\hllch  \ee
into eigenspaces of $\tauc$. The label $\ch$ takes values\vspace{-.12em}
in the character group $\Gilabels$ of the (abelian) group $\Gilabel$. 
\\Now the first crucial result of \ci4 is:\\
$\bullet$\,~The eigenspaces $\hllch$ are submodules of $\hll$, and 
they can consistently be assumed to be irreducible. This implies in particular 
that the characters of the resolved fixed points 
are the characters \bllch\ of the eigenspaces \hllch. \\
Although the definition of the eigenspaces $\hllch$ is rather implicit, we can 
still compute their characters \bllch, due to our second result:\\
$\bullet$\,~The Fourier transforms $\bllo:=\sum_\ch\ch(\JJ)\,\bllch$ 
of the characters $\bllch$ \wrt $\Gilabels$ are generalized character-valued 
indices, obtainable as the trace
  \be  \bllo(q) = \trhll(\tauc\;\qc)  \,. \labl{bllo}
$\bullet$\,~Moreover, these functions arise in the expansion 
  \be  \chil= \sum_{\LambdaP} \bllo\,\chilP  \ee
 of the {\em twining characters\/} \ci3
  \be  \chil:=\trhl(\taug\,\qg)  \ee 
of \gg\ \wrtt \ocha s $\chilP:=\trhlP(\taugP\,q_{}^{L_0'-C'/24})$ of \gP.
We therefore call them the {\em\tbf s\/} associated to $\JJ$.\\
$\bullet$\,~We can then compute the \tbf s from the \rep\ theory of \gg\ and 
\gP: the \tcha s coincide with ordinary characters of the so-called \olie s 
$\gO$ and $\gO{}'$ \cite{fusS3,furs,sfrs},
so that \bllo\ can be calculated as the ordinary
branching function $\brev b_{\brev\Lambda,\brev\Lambda'}$,
satisfying $\chiO=\sum_\LambdaOP\brev b^{}_{\brev\Lambda,\brev\Lambda'}\chiOP$,
of the coset construction $(\gO/\gO{}')_\kvo$ of \olie s.

It follows in particular that \tbf s carry a unitary \rep\ of \slz.
Their modular \smat\ $S^{\jjj}$ is the tensor product of the relevant 
S-matrices $\So$ and $\SoP$ for the \omchar s of \gg\ \resp\ \gP,
which in turn are the ordinary S-matrices of the \olie s \gO\ and $\gPO$.
By expressing the coset characters $\bllch$ as inverse Fourier transforms of
the $\bllo$, we can then determine the \smat\ $\cS$ for the coset theory \ggp. 
We find
  \be  \hsp{-.6}\Scc\Lambda\LambdaP\Mu{\Mu'} \!= \Frac{|\G|}{|\Gilabel|\cdot
  |\Gjlabel|} \hsp{-1.5}\sum_{\scriptstyle\jj\in\hsp{1.3} \atop \scriptstyle
  \hsp{1.7}\Gilabel\cap\Gjlabel}\hsp{-1.6}
  \ch^*(\JJ)\, S^{\jjj}_{\ilabel,\jlabel} \cht(\JJ) \,. \labl{Scc}
Using the modular properties
of ordinary and twining characters of \gg\ and \gP, we can show that $\cT$ -- 
which is just the T-matrix of $\g\!\oplus\!(\gP)^*$ restricted to 
orbits -- and $\cS$ generate a unitary \rep\ of \slz, and that
$\cS$ is symmetric and squares to an order-two permutation.
Finally, in all the (many) cases that we have checked explicitly, 
$\cS$ gives rise, via the Verlinde formula,
to non-negative integral fusion coefficients.

\section{\Fpr for simple current extensions}

Many non-diagonal modular invariants ${\cal Z}$ can be interpreted 
as the diagonal invariant for an extension of the chiral algebra by a group 
\Gex\ of integer spin simple currents. They can be written as
  \be  {\cal Z}=\sum_{[a]:\;Q(a)=0}  
  |\Sa| \cdot |\!\!\!\sum_{J\in\gex/\Sa}\!\!\! \X_{Ja}^{} |^2_{} , \labl z
where $\chii_a$ denotes the irreducible characters of the original 
`unextended' theory, the first sum is over all \Gex-orbits $[a]$ 
with zero monodromy charges $Q$ \cite{scya6}, and $\Sa\!\subseteq\!\gex$ is the
stabilizer $\Sa=\{J\iN\gex\,|\,Ja\eq a\}$.
As in the case of coset theories, we want to know the spectrum
of the putative conformal field theory that has this extended chiral algebra. 

For non-fixed points, i.e.\ $|\Sa|\eq1$, one can simply interpret 
$\Xt_a\eq\sum_{J\in \gex}\X_{Ja}$ as an irreducible character of the extended 
theory. Similarly,
for $|\Sa|\eq2$ or 3 the orbit $[a]$ corresponds to precisely $|\Sa|$ \irmod s
in the extended theory whose characters $\Xt$ are identical with respect to the 
unextended algebra. But for $|\Sa|\geq 4$ it can happen that the prefactor 
$|\Sa|$ or part of it must be interpreted as belonging to the extended
character rather than as a multiplicity, so that
there are as many potential interpretations of $|\Sa|$ as there are ways of 
writing it as a sum of squares. The task to determine 
the meaning of the prefactors $|\Sa|\geq 4$ (and thereby find the spectrum)
constitutes the \fpr\ problem for simple current extensions.

The main results of \ci6 are a prescription on how to interpret 
factors $|\Sa|\geq 4$ and a formula for the \smat\ $\tS$ which acts on the 
characters $\Xt$ of the extended theory.
They are obtained by solving the constraint that $\tS$ is symmetric
and together with $\tT$ (which coincides with $T$ `on orbits')
generates a unitary \rep\ of \slz, supplemented by the working hypotheses 
that the order of successive extensions does not matter
and that the characters of all resolved fixed points coming from an orbit $[a]$
are identical \wrtt unextended algebra (`fixed point homogeneity').
Since the first condition is necessary, but not sufficient for the
existence of a consistent \cft, our results are still conjectures.
Note that fixed point homogeneity means that taking (part of) $|\Sa|$
into the characters can only occur when $|\Sa|$ is a square.\,%
\futnote{Resolutions which violate fixed point homogeneity sometimes also
exist; e.g.\ the simple current invariant of $A_4$ at level 5 
has a fixed point with $|\Sa|\eq5$ for which $|\Sa|$ splits as $5=1^2+2^2$.}

Our result for the \smat\ closely resembles the formula \erf{Scc} for $\cS$:
  \be  \tS_{(a,\Psi),(b,\tilde\Psi)}=
  \sqrt{\Ia\Ib}\,\Frac{|\G|}{|\Sa|\cdot|\Sb|}\,
  \sum_{J\in U_a\cap U_b} \Psi(J)\, \So_{a,b}\, \tilde\Psi(J)^*\,. \labl0
But it involves a new ingredient, the {\em untwisted stabilizers\/} 
$U_a\!\subseteq\!\Sa$. Namely, to each $J\iN\Gex$ we associate a unitary matrix 
$\So$ which vanishes on non-fixed points of $J$ and, when restricted to fixed 
points, satisfies $(\So)^4\eq\one$ and $(\So T)^3\eq(\So)^2$, 
as well as $S^{[J^{-1}]}_{}\eq{(\So)}^{\rm t}$.
Further, for any simple current $K$ that is local with respect to $J$
one defines a complex number $\Fa KJ$ by the relation
$\So_{Ka,b}=\Fa KJ\,\eE^{2\pi\ii Q_K(b)}\So_{ab}$ between \smat\ elements.
Then $U_a:=\{J\iN\Sa\,|\, \Fa KJ\eq1 \;\;{\rm for\;all}\; K\iN\Sa\}$.

In \erf0, $\Psi$ and $\tilde\Psi$ are characters of $U_a$ and of $U_b$, \resp,
and $\Ia$ is the index $\Ia=|\Sa|/|U_a|$, which is always a square.
For the invariant \erf z this means that each fixed point $[a]$ is resolved into
$|\U_a|$ distinct fields, which are labelled by the characters of $\U_a$.
Again we checked in many cases that the formula \erf0 leads to non-negative 
integral fusion coefficients.

For WZW models, the matrices $\So$ can be identified (up to a known over-all 
phase) with the modular S-matrices of the associated orbit Lie algebras.
For other \cfts\ we must postulate the existence of the relevant matrices $\So$.
We conjecture that they do not constitute a new structure, but rather 
coincide with the matrices that describe the modular S-transformation on the 
space \block\ of chiral one-point blocks on the torus with 
insertion of the relevant simple current $J(z)$.
Indeed, our $\So$ share all known properties of these matrices, including
certain recently deduced \cite{bave} trace relations.
Moreover, in \cite{bant6} it has been shown that the untwisted stabilizers
appear naturally in the framework of one-point blocks, where they possess
a cohomological interpretation related to a basis choice for 
the space \block\ of one-point blocks.

\bigskip\bigskip \small\noindent{\bf Acknowledgement.}
J.\ Fuchs was supported by a Heisenberg fellowship of the Deutsche
For\-schungs\-gemeinschaft. C.\ Schweigert
was supported in part by the Director, Office of Energy Research,
Office of Basic Energy Sciences, of the U.S.\ Department of Energy under
Contract DE-AC03-76F00098, and in part by the National Science Foundation
under grant PHY95-14797.

 \newcommand\wb       {\,\linebreak[0]} \def\wB {$\,$\wb}
 \newcommand\Bi[1]    {\bibitem{#1}}
 \newcommand\J[5]     {\ {\sl #5}, #1 #2 ({#3}) {#4} }
 \newcommand\Prep[2]  {{\sl #2}, preprint {#1}}
 \def\jf    {J.\ Fuchs}
 \def\bams  {Bull.\wb Amer.\wb Math.\wb Soc.}
 \def\fuaa  {Funct.\wb Anal.\wb Appl.}
 \def\ijmp  {Int.\wb J.\wb Mod.\wb Phys.\ A}
 \def\nupb  {Nucl.\wb Phys.\ B}
 \def\phlb  {Phys.\wb Lett.\ B}
 \def\phrd  {Phys.\wb Rev.\ D}
 \def\comp  {Com\-mun.\wb Math.\wb Phys.}
 \def\A       {Algebra}
 \def\alg     {algebra}
 \def\Infdim  {Infinite-dimensional}
 \def\Q       {Quantum\ }
 \def\Rep     {Representation}
 \def\syms    {sym\-me\-tries}
 \def\wzw     {WZW\ }
 \version\versionno 
\end{document}